\documentclass[a4paper,11pt]{article}

\usepackage[a4paper,left=2.73cm,right=2.7cm,top=3cm,bottom=3.5cm]{geometry}
\usepackage{amsmath,amssymb,graphicx,colortbl,xcolor,caption,subcaption,wrapfig}
\usepackage[colorlinks=true,linktocpage=true,linkcolor=blue,citecolor=blue]{hyperref}

\usepackage[all]{xy}
\definecolor{gray}{rgb}{.9,.9,.9}

\newcommand{\dd}{\mathrm{d}}

\numberwithin{equation}{section}

\newcommand{\eqq}[1]{(\ref{#1})}

\newcommand{\mt}[1]{\textrm{\tiny #1}}
\newcommand{\nc}{N}
\newcommand{\nf}{N_f}
\newcommand{\gym}{g_\mt{YM}}

\newcommand{\sac}{\, , \qquad}

\newcommand{\be}{\begin{equation}}
\newcommand{\ee}{\end{equation}}
\newcommand{\bal}{\begin{align}}
\newcommand{\bse}{\begin{subequations}}
\newcommand{\ese}{\end{subequations}}
\newcommand{\bea}{\begin{eqnarray}}
\newcommand{\eea}{\end{eqnarray}}

\begin{document}

\begin{titlepage}

\hfill{ICCUB-15-015}

\vspace{1cm}
\begin{center}

{\LARGE{\bf  Three-dimensional super Yang--Mills \\[3mm] 
	     with unquenched flavor}}

\vskip 45pt
{\large \bf Ant\'on F. Faedo$^{1}$, David Mateos$^{1,2}$ and Javier Tarr\'\i o$^{1}$}

\vskip 10pt
{$^{1}$Departament de F\'\i sica Fonamental and Institut de Ci\`encies del Cosmos, \\ Universitat de Barcelona, Mart\'\i\  i Franqu\`es 1, ES-08028, Barcelona, Spain.}\\

\vskip 10pt
{$^{2}$Instituci\'o Catalana de Recerca i Estudis Avan\c cats (ICREA), \\
Passeig Llu\'\i s Companys 23, ES-08010, Barcelona, Spain}

\vskip 10pt
%{e-mails: xxx}
\end{center}

\vspace{10pt}
\abstract{\normalsize
We construct analytically the  gravity duals of three-dimensional, super Yang--Mills-type theories with $\mathcal N=1$ supersymmetry coupled to $\nf$ quark flavors. The backreaction of the quarks on the color degrees of freedom is included, and corresponds on the gravity side to the  backreaction of $\nf$ D6-branes on the background of $\nc$ D2-branes. The D6-branes are smeared over the compact part of the geometry, which must be a six-dimensional nearly K\"ahler manifold in order to preserve supersymmetry. For massless quarks, the solutions flow in the IR to an $AdS_4$ fixed point dual to a Chern--Simons-matter theory. For light quarks the theories exhibit quasi-conformal dynamics (walking) at energy scales $m_q \ll E \ll \lambda \nf / \nc$, with $\lambda=\gym^2 \nc$ the 't Hooft coupling.

}

\end{titlepage}

\tableofcontents

\hrulefill
\vspace{15pt}

\section{Introduction} 
%%%%%%%%%%%%%%%%%%%%%%%%%%%%%%%%%%%%%%%%%%%%%%%%%%%%%%%%%%%%%%
Three-dimensional, ${\cal N}=8$ super Yang--Mills (SYM) theory with gauge group $SU(\nc)$ was among the first detailed examples of holographic duality \cite{Itzhaki}. In three dimensions the gauge coupling $\gym^2$ has dimensions of energy and thus the theory possesses a non-trivial renormalization group (RG) flow. Since the theory is asymptotically free, it provides an example in which the ultraviolet (UV) is nicely behaved. The holographic description of this theory is obtained from the supergravity solution sourced by $\nc$ D2-branes in flat space. The amount of supersymmetry can be easily reduced to ${\cal N}=1$ by replacing the flat space transverse to the D2-branes, which is a cone over $S^6$, by a $G_2$-holonomy cone whose base must be a six-dimensional nearly K\"ahler (NK) manifold ${\cal M}_6$ different from the six-sphere \cite{Acharya2}. In this case the dual gauge theory is presumably a quiver gauge theory. 

Our goal in this paper will be to understand the effect of adding $\nf$ flavors of  fundamental matter to the theories above.  In all cases the flavored theory will be $\mathcal N=1$ supersymmetric. We will loosely refer to this matter as `quarks' despite the fact that it will include both bosonic and fermionic degrees of freedom. We will see that their inclusion leads to interesting infrared (IR) dynamics such as the appearance of a Chern--Simons (CS) matter theory in the case of massless quarks or of `walking' (quasi-conformal) dynamics in the case of light quarks.

The addition of quarks on the gauge theory side  corresponds to the addition of D6-branes on the gravity side \cite{Karch}.  We will work with unquenched quarks, meaning that their backreaction on the adjoint (color) degrees of freedom is included. On the gravity side we will therefore include the backreaction of the D6-branes on the D2-brane geometry. 

The case when the internal manifold is $S^6$ and all the D6-branes are overlapping was studied in \cite{Pelc:1999ms,Cherkis}. The supersymmetric solutions constructed in these references depend on two radial coordinates: the radius transverse to the D2-branes but parallel to the D6-branes, and the radius transverse to both the D6- and the D2-branes. Despite their beauty, these solutions are difficult to generalize because the dependence on two radii translates into the requirement to solve non-linear partial differential equations. For example, it has not been possible to find the generalizations corresponding to the introduction of a non-zero temperature or a non-zero charge density into the system.\footnote{See however \cite{GomezReino:2004pw} for a perturbative, finite-temperature solution near the core of the D6-branes, and \cite{Erdmenger:2004dk} for a computation of the meson spectrum in the same near-core limit.}
 
In order to reduce the supergravity equations to ordinary differential equations, we will smear the D6-branes over the internal manifold \cite{Bigazzi} (see  \cite{Nunez:2010sf} for a review of this approach). This, together with the reduction to first-order equations implied by supersymmetry, will allow us to construct the solution essentially analytically. 

The presence of the quarks  leaves the UV properties of the gauge theory unmodified, but it  changes the IR dynamics. In the case of massless quarks, the solution flows to an $AdS_4$ fixed point in the IR. We will argue that, in general, the dual gauge description is a CS-matter theory. In the particular case in which the NK  manifold is $\mathbb{CP}^3$ we will be able to identify this theory as the flavored version \cite{Conde} of the Ooguri-Park solution \cite{Ooguri}, which is itself an $\mathcal{N}=1$ deformation of the ABJM \cite{Aharony} theory.
In the case of  quarks that are light compared to the scale $\lambda \nf/\nc$ set by the 't Hooft coupling $\lambda=\gym^2 \nc$, the theory exhibits `walking' or quasi-conformal dynamics in the energy range $m_q \ll E \ll \lambda \nf / \nc$.

%%%%%%%%%%%%%%%%%%%%%%%%%%%%%%%%%%%%%%%%%%%%%%%%%%%%%%%%%%%%%%
\section{Flavorless solutions and nearly K\"ahler manifolds}
%%%%%%%%%%%%%%%%%%%%%%%%%%%%%%%%%%%%%%%%%%%%%%%%%%%%%%%%%%%%%%

Since three-dimensional SYM-type theories are asymptotically free, in the UV a perturbative description is possible. At an energy scale $E\sim \lambda$ the theories becomes strongly coupled. If  $\nc$ is large, then this regime can be described holographically by the gravitational background sourced by $\nc$ D2-branes; in terms of the string coupling and the string length, the Yang--Mills coupling is 
\be
\gym^2=g_s/\ell_s \,.
\label{coupling}
\ee
If the space transverse to the D2-branes is flat space then maximal, 
$\mathcal{N}=8$ supersymmetry is preserved (in three-dimensional language) and the gauge theory is $SU(N)$ SYM. In order to decrease the amount of supersymmetry  --- as the addition of flavor will do anyhow --- the stack of branes has to be positioned at the tip of a Ricci-flat cone with reduced holonomy \cite{Acharya2}. This replaces the transverse flat space, which in polar coordinates is a cone over the six-sphere $S^6$. In order to preserve $\mathcal{N}=1$ supersymmetry the cone must have $G_2$ holonomy. The base of a $G_2$-cone is a nearly-K\"ahler manifold. Since properties of NK manifolds will play an important role, we now proceed to review them.

%%%%%%%%%%%%%%%%%%%%%%%%%%%%%%%%%%%%%%%%%%%%%%%%%%%%%%%%%%%%%%
%\subsection{The internal geometry}\label{NK}
%%%%%%%%%%%%%%%%%%%%%%%%%%%%%%%%%%%%%%%%%%%%%%%%%%%%%%%%%%%%%%

NK geometries appear naturally in the classification of almost Hermitian manifolds \cite{Gray}. A NK manifold is an almost Hermitian manifold whose fundamental form $J$ satisfies the condition
\begin{equation}
3\,\nabla J\,=\,\dd J\,,
\end{equation}
with $\nabla$ the derivative associated to the Levi--Civita connection. In this terminology, the K\"ahler condition is $\nabla J=0$. Powerful splitting theorems reduce the study of these geometries to six dimensions \cite{Nagy}, and we will restrict ourselves to this case in the following. The importance of these geometries in physics (and in particular in string-theory) stems from the fact that they admit Killing spinors. Indeed, in six dimensions, a manifold admits a Killing spinor if and only if it is NK \cite{Grunewald}. As a consequence, such a manifold is Einstein with positive curvature.
 
This property makes NK manifolds suitable geometries  to support supergravity solutions preserving some amount of supersymmetry. In fact, an equivalent way of defining a NK manifold that makes this more apparent is the following. The Riemannian cone
\begin{equation}\label{cone}
\dd s^2\left(\mathcal{C}_7\right)\,=\,\dd r^2+r^2\,\dd s^2\left(\mathcal{M}_6\right)
\end{equation}
has $G_2$ holonomy if and only if $\mathcal{M}_6$ is NK \cite{Bar}. As is well known, $G_2$ holonomy is the condition to preserve minimal supersymmetry in four dimensions starting from eleven-dimensional M-theory, in analogy with the $SU(3)$ holonomy of Calabi--Yau manifolds that is required when starting from ten-dimensional string theory. 
 In particular, models with chiral fermions can be obtained considering M-theory in the presence of $G_2$-cone singularities like (\ref{cone}), as shown in \cite{Atiyah, Acharya}.

These internal geometries also appeared in the quest for stabilizing moduli in string compactifications to Minkowski (and related)  vacua. In \cite{Behrndt} it was found that massive type IIA supergravity admits $\mathcal{N}=1$, ${ AdS}_4\times\mathcal{M}_6$ solutions with $\mathcal{M}_6$ being NK. In the context of flux compactifications, the most suitable language to analyze the backgrounds is that of $G$-structures. This leads us to another way of characterizing NK geometries, i.e.~as six-dimensional manifolds admitting an $SU(3)$ structure with only $\mathcal{W}_1$ non-vanishing among the torsion classes in the codification of \cite{Chiossi}. Possessing $SU(3)$ structure implies the existence of a globally defined, real two-form $J$ (associated to the almost complex structure) together with a globally defined, complex three-form $\Omega$, satisfying
\begin{eqnarray}\label{$SU(3)$structure}
J\wedge\Omega&=&0\,,\nonumber\\[2mm]
\frac13\,J\wedge J\wedge J&=&\frac{i}{4}\,\Omega\wedge\overline\Omega\,.
\end{eqnarray}
From the class of $SU(3)$ structure manifolds, nearly-K\"ahlerness is selected by the requirements
\begin{eqnarray}\label{NKness}
\dd J&=&\frac32\,{\rm Im}\left[\,\overline{\mathcal{W}}_1\,\Omega\,\right]\,,\nonumber\\[2mm]
\dd\Omega&=&\mathcal{W}_1\,J\wedge J\,,
\end{eqnarray}
that is, the torsion classes $\mathcal{W}_2,\dots,\mathcal{W}_5$ corresponding to other $SU(3)$ representations that could occur in \eqq{NKness} all vanish. For the purposes of this paper, equations (\ref{$SU(3)$structure}) and (\ref{NKness}) will be the defining properties of the internal geometries. 

Considerably less attention has been paid to NK manifolds in the context of the gauge/string duality, despite the early observation in \cite{Acharya2} that they emerge as the transverse space to D2-branes preserving $\mathcal{N}=1$ supersymmetry and generalizing $S^6$. In this respect, a parallel can be traced with D3-branes in type IIB supergravity and Sasaki--Einstein (SE) manifolds. Just like a stack of D3-branes at the tip of a cone over a SE manifold is dual to a  gauge theory with minimal supersymmetry in four dimensions, a stack of D2-branes at the tip of a $G_2$-cone as in (\ref{cone}) is dual to a gauge theory with minimal supersymmetry in three dimensions. Despite this resemblance, there are also  some important differences. First, the D2-brane near horizon geometry is not $AdS$, as opposed to the D3-brane case. This means that one cannot use the familiar tools of conformal field theories (CFT) to analyze the dual gauge theory. And second, compared with Sasakian geometry, NK manifolds are still not well understood from the mathematical viewpoint, though several important results are known (for reviews see \cite{Verbitsky,Boyer}).   

The comparison between NK manifolds in six dimensions and SE manifolds in five dimensions is summarized in the following table:

\begin{table}[h]
\resizebox{0.95\textwidth}{!}{\begin{minipage}{\textwidth}
\begin{tabular}{cccccccccc}
\hline
$\mathcal{M}_d$  &\phantom{sp} &$\mathcal{C}\left(\mathcal{M}_d\right)$&\phantom{sp}&Gauge theory dual&\phantom{sp}&$G$-structure&\phantom{sp}&Globally defined forms
\\
\hline
%\rowcolor{gray}	
NK&&$G_2$-cone&&$D=3$,\,\,$\mathcal{N}=1$\,\,SYM&&$SU(3)$&&$J_{(2)}\,,\,\,\Omega_{(3)}\,$\\
SE&&Calabi--Yau&&$D=4$,\,\,$\mathcal{N}=1$\,\,SCFT&&$SU(2)$&&$\eta_{(1)}\,,\,\,J_{(2)},\,\,\Omega_{(2)}\,$\\
\hline
\end{tabular}
\caption{Some key properties of NK manifolds in comparison with SE manifolds. $J_{(2)}$ always denotes a real two-form. $\Omega_{(n)}$ is a complex $n$-form. $\eta_{(1)}$ is a real one-form.}
\label{NKvsSE}
\end{minipage} }
\end{table}

The complete list of known, regular, compact, six-dimensional NK  manifolds is as follows:\footnote{Besides, there are infinite families of examples with orbifold or conical singularities \cite{Cvetic, Behrndt2, Fernandez}.}
\begin{eqnarray}\label{NKlist}
{S}^6&\simeq&\frac{{ G}_2}{ SU(3)}\nonumber\\[4mm]
\mathbb{CP}^3&\simeq&\frac{ Sp(2)}{{ SU(2)}\times{ U(1)}}\nonumber\\[4mm]
{ S}^3\times{ S}^3&\simeq&\frac{{ SU(2)}\times{ SU(2)}\times{SU(2)}}{SU(2)}
\nonumber\\[4mm]
{\rm F}(1,2)&\simeq&\frac{SU(3)}{{ U(1)}\times{ U(1)}}
\end{eqnarray}
All these manifolds are homogeneous. Moreover, any six-dimensional, homogeneous, NK manifold is isometric to one of them \cite{Butruille}. We emphasize that the metric on the $\mathbb{CP}^3$ that is compatible with the NK structure is not the perhaps-more-familiar Fubini--Study metric used in the   ABJM construction \cite{Aharony}, which is instead K\"ahler. Regarding  
$\mathbb{CP}^3$ as an $S^2$ fibration over $S^4$, the NK metric is squashed with respect to the K\"ahler one. As a consequence, the isometry is reduced to $Sp(2)\simeq SO(5)\subset SU(4)$. In a similar manner, the NK metric on ${S}^3\times{ S}^3$ is not the product of the round metrics,
 %but the spheres are at a non-vanishing angle, 
 so again the isometry is reduced. For more details about these cosets we refer the reader to \cite{Atiyah}. 
%%%%%%%%%%%%%%%%%%%%%%%%%%%%%%%%%%%%%%%%%%%%%%%%%%%%%%%%%%%%%%
\section{Adding flavor}\label{setup}
%%%%%%%%%%%%%%%%%%%%%%%%%%%%%%%%%%%%%%%%%%%%%%%%%%%%%%%%%%%%%%
\subsection{Generalities}

We have argued that a stack of $\nc$ D2-branes placed at the tip of the $G_2$-cone (\ref{cone}) provides the holographic dual --- at least in some energy range --- to a three-dimensional gauge theory with minimal supersymmetry, that is, two real supercharges. When the internal NK geometry is taken to be the six-sphere, there is an enhancement of supersymmetry to $\mathcal{N}=8$ and the resulting duality was examined in \cite{Itzhaki}. Still largely unexplored are the detailed field theory duals for the remaining cosets in the list (\ref{NKlist}), as well as for other putative NK manifolds to be found.\footnote{See \cite{Loewy} for a proposal for the quiver theory dual to the NK metric on $\mathbb{CP}^3$.} 

Setting aside what is the specific SYM dual to these gravitational solutions, we will focus instead on the inclusion of fundamental matter, to which we will also refer as `flavor' or `quarks'. This is achieved by adding a new set  of branes to the gravitational system \cite{Karch}. Since we wish to preserve supersymmetry and we want the fundamental degrees of freedom to propagate along the 2+1 gauge theory directions (see e.g.~the discussion in Section 5.5 of \cite{CasalderreySolana:2011us}), we will introduce an additional stack of $\nf$ D6-branes. The relative orientation between the color and the flavor branes is indicated by the following array:
\begin{center}
\begin{tabular}{lccccccccc}\hline
  & $x^1$ & $x^2$ & $r$ & \multicolumn{6}{c}{${\rm NK}$} \\
\hline
%\rowcolor{gray}
D2 &  $\times$ & $\times$ & $\cdot$ & $\cdot$ & $\cdot$ & $\cdot$ & $\cdot$ & $\cdot$ & $\cdot$ \\
D6 & $\times$ & $\times$ & $\times$ & $\times$ & $\times$ & $\times$ & $\cdot$ & $\cdot$ & $\cdot$\\
\hline
\end{tabular}
\label{aa}
\end{center}
with each D6-brane wrapping a three-dimensional submanifold  inside the NK  space transverse to the D2-branes. 

The case when all the D6-branes wrap the same submanifold  and the internal manifold is an $S^6$  was studied in \cite{Pelc:1999ms,Cherkis} and generalized to other manifolds in \cite{Acharya:2003ii}.
As explained in the Introduction, the supersymmetric solutions constructed in those references solve partial differential equations and are therefore difficult to generalize. 
This technical difficulty can be overcome by smearing the flavor branes appropriately \cite{Bigazzi}. Whenever there is a large number of branes, it is possible to distribute them along their transverse directions --- the dots in the row of the D6-branes in the array above --- in such a way that the dependence on the second radius disappears. In practice one has traded the partial differential equations for ordinary ones.
 
 In our case we will distribute the D6-branes in a way that preserves the $SU(3)$ structure of the NK internal geometry. We will see that this requirement essentially fixes the distribution of D6-branes uniquely, and moreover that it is compatible with preservation of  supersymmetry.

The D6-branes will contribute to the energy-momentum tensor and thus modify the metric  originally sourced only by the D2-branes. The parameter that controls the relative influence of fundamental matter with respect to the initial color branes is \cite{Nunez:2010sf}  
\begin{equation}\label{backparameter}
\frac{N_f}{\nc}\,\,g_{\rm eff}^2\,.
\end{equation}
The effective dimensionless  coupling $g_{\rm eff}$ is defined as in \cite{Itzhaki}
\begin{equation}\label{geff}
g_{\rm eff}^2\,=\,\gym^2\,\nc\,U^{-1}\,=\,\lambda\,U^{-1}\,,
\end{equation}
where $U$ is a radial coordinate on the gravity side that is dual to an energy scale in the gauge theory. The so-called `probe approximation' on the gravity side, in which the backreaction of the flavor branes into the geometry is neglected, is justified when  $\nf \, g_{\rm eff}^2 / \nc$ is small. On the field theory side this corresponds to a `quenched' approximation in which the flavor degrees of freedom are treated as probes of the gluon-plus-adjoint-matter-dominated dynamics. If $\nf \, g_{\rm eff}^2 / \nc$ is not small, then the backreaction of the flavor branes on the geometry must be included. In the gauge theory this corresponds to a situation with `unquenched' degrees of freedom in the fundamental representation. Note that, in the context of the large-$\nc$ expansion, a necessary condition for the fundamental matter to be unquenched is that the limit $\nc \to \infty$ is taken in such a way that $\nf /\nc \neq 0$. This way of taking the large-$\nc$ limit is usually referred to as the Veneziano limit. 

It is clear from (\ref{backparameter}) and (\ref{geff}) that, for fixed $\lambda$ and non-vanishing  $N_f/\nc$, there is an energy scale, or equivalently a radial position in the geometry, at which $\nf \, g_{\rm eff}^2 / \nc$ becomes of order unity. Below this scale the backreaction of the D6-branes must be included and we expect the geometry to be significantly modified. 

In the opposite limit, for high enough energies, the backreaction of the flavor branes decreases. We therefore expect that the solution of the color branes, dual to pure super Yang--Mills, will be recovered in the UV, which is in agreement with the theory being superrenormalizable. As we will see, all these expectations will be confirmed by our explicit solutions. 

%%%%%%%%%%%%%%%%%%%%%%%%%%%%%%%%%%%%%%%%%%%%%%%%%%%%%%%%
\subsection{Ansatz}

We will now write down the ansatz for the supergravity fields in our solution and derive the corresponding BPS equations, which we will solve in subsequent sections. As indicated by the D2/D6 array above, the ten-dimensional geometry consists of three Minkowski directions, a radial coordinate and a six-dimensional internal manifold, which is assumed to possess   an $SU(3)$ structure. This implies that there exist a real two-form $J$ and a complex three-form $\Omega$ defining the structure and subject to the compatibility conditions (\ref{$SU(3)$structure}). Given an appropriate set of vielbeins $e^a$, $a=1,\dots,6$, at least locally we can write
\begin{equation}\label{local$SU(3)$}
\begin{array}{rcl}
J&=&e^{12}+e^{34}+e^{56}\,,\\[4mm]
\Omega&=&\left(e^1+i\,e^2\right)\wedge\left(e^3+i\,e^4\right)\wedge\left(e^5+i\,e^6\right)\,.
\end{array}
\end{equation}
If in addition the manifold is NK, as we assume, the differential relations (\ref{NKness}) are verified. Without loss of generality, the torsion class can be taken to be real and adjusted to $\mathcal{W}_1=2$, so we have the differential conditions
\begin{equation}\label{Torcond}
\dd J\,=\,3\,{\rm Im}\,\Omega\,,\qquad\qquad\qquad\dd{\rm Re}\,\Omega\,=\,2\,J\wedge J\,,
\end{equation}
together with the Hodge duals with respect to the internal metric
\begin{equation}
*_6J\,=\,\frac12J\wedge J\,,\qquad\qquad\qquad*_6\Omega\,=\,-i\,\Omega\,.
\end{equation}
The vast simplification implied by these assumptions follows from the fact that we have a natural set of forms to employ in the ansatz for the supergravity fields as well as the possibility to preserve supersymmetry.  

Let us begin by writing down the ansatz for the supergravity forms in the Ramond-Ramond (RR) sector. Since the solution is sourced by D2- and D6-branes, we expect the RR six- and two-form field strengths to be non-zero. As explained in Appendix \ref{details}, the distribution of backreacting D6-branes gives rise to a violation of the Bianchi identity for the RR two-form,
\begin{equation}
\dd F_2\,=\,-2\kappa^2\,T_{\rm D6}\,\Xi\,,
\label{bianchi}
\end{equation}
where $\Xi$ is the so-called smearing form that indicates how the flavor branes are distributed in the internal directions. Intuitively, this is simple to understand. The D6-branes couple minimally to the RR potential $C_7$, and hence their presence leads to a source on the right-hand side of the equation of motion for its field strength, i.e.~we have $\dd * F_8 \sim \Xi$, where $\Xi$ measures the local density and the orientation of D6-branes at any given point. Since by definition $F_2 = - * F_8$, this leads to eqn.~\eqq{bianchi}.

Given the forms at our disposal, the simplest ansatz for the RR fluxes reads
\begin{eqnarray}
\label{RRansatz}  F_2&=&Q_f\,J \,,\\[4mm]
\label{RRansatz2} F_6&=&\frac{Q_c}{6}\,J\wedge J\wedge J\,.
\end{eqnarray}
As we will see below, for massless quarks the fact that $J$ is not closed is crucial for the consistency of this ansatz with the violation of the Bianchi identity. As usual, the six-form sourced by the D2-branes is proportional to the volume form of the internal space. The parameters $Q_c$ and $Q_f$ have dimensions of (length)$^5$ and (length)$^1$ and are related to the number of D2- and D6-branes, respectively, or equivalently to the rank of the gauge group and the number of flavors, through the quantization condition
\begin{equation}
\frac{1}{2\kappa^2T_{\text{D}p}}\,\int\,F_{8-p}\,=\,\frac{1}{(2\pi\ell_s)^{7-p}g_s}\,\int\,F_{8-p}\,=\,N_p\,.
\end{equation}
Using \eqq{RRansatz} and \eqq{RRansatz2} this immediately gives
\begin{equation}\label{quantcond}
Q_c\,=\,\frac{(2\pi\ell_s)^{5}g_s}{V_6}\,\nc\,,\qquad\qquad Q_f\,=\,\frac{(2\pi\ell_s)g_s}{V_2}\,N_f\,,
\end{equation}
where the dimensionless quantities $V_6$ and $V_2$ are the volume of the internal manifold and $\int J$, respectively. The computation of $V_2$  requires the knowledge of an explicit realization of the NK structure. We emphasize that, in general, $Q_f=Q_f(r)$ can be a function of the radial coordinate, as will be the case for massive quarks. Note that, up to numerical coefficients, we have that 
\be
Q_c \sim \lambda \, \ell_s^6 \sac Q_f \sim \lambda \frac{\nf}{N} \, \ell_s^2\,.
\ee
As always in the context of the gauge/string correspondence, the powers of $\ell_s$ 
will cancel out in the computation of gauge theory observables.

It is instructive at this point to come back to the analogy with the more familiar SE geometries transverse to D3-branes. In this case the flavored solution was found in \cite{Benini} and contains an additional stack of D7-branes, whose backreaction induces a violation of the Bianchi identity for the RR one-form: 
\begin{equation}
\label{BianchiF1}
\dd F_1\,=\,-2\kappa^2\,T_{\rm D7}\,{\Xi}\,.
\end{equation}
As shown in Table \ref{NKvsSE}, in a SE geometry we have at our disposal a globally defined, non-closed one-form. The natural ansatz for the one-form field strength used in \cite{Benini} was thus
\begin{equation}
F_1\,=\,{Q}_f\,\eta\,.
\label{F1}
\end{equation}
Equations \eqq{BianchiF1} and \eqq{F1} are the SE analog of our NK equations \eqq{bianchi} and \eqq{RRansatz}.

Equations (\ref{Torcond}), \eqq{bianchi} and \eqq{RRansatz} determine the distribution of the flavor branes, i.e.~$\Xi$, up to an arbitrary  function $Q_f (r)$ as
\begin{equation}
\Xi\,=\,-\frac{1}{2\kappa^2\,T_{\rm D6}}\Big( 
Q_f'\,\dd r\wedge J+3\,Q_f\,{\rm Im}\,\Omega\Big)\,,
\end{equation}
where the $'$ denotes differentiation with respect to the radial coordinate $r$. We will see that for massless quarks $Q_f$ is just a constant, meaning that in this case the branes are smeared purely along the directions spanned by ${\rm Im}\,\Omega$.

We now turn to the Neveau-Schwarz (NS)  sector of the theory. Working in string frame, we choose to parameterize it  as
\begin{eqnarray}\label{NSansatz}
\dd s_{\rm s}^2&=&h^{-\frac12}\,\dd x^2_{1,2}+h^{\frac12}\,e^{2\chi}\,
\dd s^2\left(\mathcal{C}_7\right) \,,\nonumber\\[4mm]
\dd s^2\left(\mathcal{C}_7\right) &=& \dd r^2+r^2\,\dd s_6^2 \,,\nonumber\\[4mm]
e^{\Phi}&=&h^{\frac14}\,e^{3\chi} \,,\nonumber\\[4mm]
 H &=& 0 \,.
\end{eqnarray}
The internal metric $\dd s_6^2=e^a\,\delta_{ab}\,e^b$ is that of a NK manifold, associated to the vielbeins introduced in \eqq{local$SU(3)$}  and normalized so that its Ricci tensor satisfies $R_{ab}=5\,\delta_{ab}$. The $G_2$-cone metric (\ref{cone}) thus appears explicitly in our ansatz. The entire system is specified by two  functions of the radial coordinate, $h(r)$ and $\chi(r)$, corresponding roughly to the dilaton and the breathing mode  (the volume modulus of the internal manifold). Even without allowing for deformations of the internal NK manifold, one may have expected the most general ansatz preserving the required symmetries to contain an additional function, but this can always be integrated in terms of the other two --- see Appendix \ref{details} for details. The virtue of the parameterization \eqq{NSansatz} is that the flavorless solution is straightforwardly recovered by setting $N_f=\chi=0$ and taking $h$ to be the usual harmonic function of the D2-branes, $h=Q_c/5r^5$.

At first sight, it may seem overly restrictive to not allow for any deformation of the internal metric, specially taking into account that some form of squashing is present in every solution with smeared flavor branes discussed in e.g.~\cite{Nunez:2010sf}. At the technical level, this difference is due to the fact that in our case there is no natural way of writing a generic NK manifold as a fiber over some base. In contrast, every SE manifold is a $U(1)$ fiber over a K\"ahler--Einstein base, and the backreaction of the flavor branes produces  a squashing between the fiber and the base. 

One immediate consequence of this difference is that the metrics of our flavored solutions possess  the same isometries as the unflavored ones. Moreover, when the NK manifold is not the six-sphere, the almost complex structure $J$, and therefore the fluxes in our ansatz, preserve all the isometries of the metric \cite{Butruille}. It thus follows  that the gauge theories with fundamental matter dual to our solutions will be invariant under  the same global symmetries as the corresponding unflavored ones, the exception being the maximally supersymmetric case dual to $S^6$. In this case  the metric enjoys a full $SO(7)$ invariance but this is broken down to $G_2\subset{ SO(7)}$ by the two-form flux (\ref{RRansatz}). This $G_2$ subgroup can easily be understood as  the numerator of the coset $S^6\simeq\,\,$$G_2$/$SU(3)$.\footnote{This $G_2$ subgroup should not to be confused with the holonomy of the cone (\ref{cone}). The cone over a six-dimensional NK manifold always has $G_2$ holonomy, but the isometry group of the NK manifold may be different. For example, as mentioned above for $\mathbb{CP}^3$ viewed as a NK manifold the isometry group is $Sp(2) \simeq SO(5)$.}

That this is the preserved symmetry follows for instance from the observation in \cite{Cassani} that the forms of the six-sphere NK structure, $J$ and $\Omega$, coincide with the $G_2$ left-invariant forms on the coset. The fact that the fluxes preserve a smaller amount of symmetry than the metric is consistent with the equations of motion, because the fluxes enter the stress tensor that sources the metric only quadratically, and the square of the fluxes is more symmetric than the fluxes themselves. 

\subsection{BPS equations}
Since we are seeking supersymmetric solutions, our next task is to write down the corresponding BPS equations that the  functions in our ansatz must obey. Typically this would require studying the  fermionic variations of the  supergravity fields, but in our case we can bypass this by making use of some results in the literature, together with the mathematical properties of the background. Indeed, the ingredients needed to write the BPS equations are contained in \cite{Haack}. This reference studies  the construction of four-dimensional domain walls in type IIA supergravity, understood as solutions possessing Poincar\'e symmetry in three dimensions. Minimal supersymmetry is also imposed. NK manifolds are particularly simple examples of internal geometries potentially preserving $\mathcal{N}=1$, so they are considered in detail. Our ansatz verifies all the assumptions in \cite{Haack} except for the violation of the Bianchi identity for the two-form. Adapting the results of \cite{Haack} to account for this fact, we conclude that the BPS equations in our case read
\begin{eqnarray}
\label{bps}
\chi'&=&\frac{Q_f}{r^2}\,e^{2\chi}\,,\nonumber\\[4mm]
h'&=&-\frac{Q_c}{r^6}\,e^{-2\chi}-\frac{3\,Q_f}{r^2}\,e^{2\chi}\,h\,.
\end{eqnarray}

Since each of the D6-branes in our solution preserves supersymmetry, we expect that each of them wraps a calibrated four-cycle inside the $G_2$-cone transverse to the D2-branes. The fact that this cone possesses $G_2$-holonomy and not just a $G_2$-structure (i.e.~that the $G_2$-structure is torsion-free, as implied by the fact that the base is NK) means that the corresponding associative three-form $\Psi$ and co-associative four-form $\tilde \Psi = *_7\Psi$ are both closed, and in fact they are both calibration forms. In terms of the NK forms and the radial coordinate on the cone, the co-associative form is given by 
\be
\tilde \Psi = r^3\,\dd r\wedge{\rm Re}\,\Omega+\frac12\,r^4\,J\wedge J \,.
\ee
We thus expect that the full world volume of a given D6-brane obeys a generalized calibration condition of the form \cite{Gutowski:1999tu}
\begin{equation}\label{calcond}
\dd\left(e^{-\Phi}\,\mathcal{K}\right)\,=\,*F_2\,=\,F_8\,,
\end{equation}
with $\mathcal{K}$ a generalized calibration form proportional to $\tilde \Psi$. We have checked by direct calculation that, upon using the BPS equations, this condition is indeed obeyed by 
\be
\mathcal{K}=h^{\frac14}\,e^{4\chi}\,\dd^3x\wedge \tilde \Psi \,.
\ee
By definition of a calibration, it then follows that the world volume action of a single D6-brane can be written as 
\begin{equation}
S_{D6}\,=\,-T_{\rm D6}\,\int\,\left(e^{-\Phi}\,\mathcal{K}-C_7\right)\,,
\end{equation} 
where $\dd C_7=F_8$ and pull-backs onto the brane's worlvolume are understood.  The general results in \cite{Koerber:2007hd} imply that, in the presence of callibrated branes, the sourced-modified equations of motion are implied once the (violated) Bianchi identities are imposed. We have checked that this is the case for all the equations of motion listed in Appendix \ref{details}.

%%%%%%%%%%%%%%%%%%%%%%%%%%%%%%%%%%%%%%%%%%%%%%%%%%%%%%%%%%%%%%
\section{Massless quarks and an infrared fixed point}
\label{masslessquarks}

We begin by considering the solution for massless quarks which, as we will justify in the next section, corresponds to setting $Q_f$ to a constant. The system \eqq{bps} of first order BPS equations is easily integrable. The general solution, in terms of  two integration constants $c_\chi$ and $c_h$, is
\begin{eqnarray}\label{massless}
e^{-2\chi}&=&c_\chi+\frac{2\,Q_f}{r}\,,\nonumber\\[4mm]
h&=&\frac{\left(2Q_f+c_\chi \,r\right)^2}{r^6}\,\left[\frac{Q_c}{315\,Q_f^5}\,\left(35Q_f^4-20Q_f^3c_\chi\,r+12Q_f^2c_\chi^2\,r^2-8Q_fc_\chi^3\,r^3+8c_\chi^4\,r^4\right)\right.\nonumber\\
&&+\left.\left(c_h-\frac{8Q_c}{315Q_f^5}\right)\,\left(\frac{c_\chi^9\,r^9}{2Q_f+c_\chi\,r}\right)^\frac12\right]\,.
\end{eqnarray}
Without loss of generality we will set $c_\chi=1$, since this can be achieved through the rescalings 
\be
x^\mu \to c_\chi^3 \,x^\mu \sac r \to c_\chi^{-1} \, r \,. 
\ee
The interpretation of $c_h$ is clarified by examining the solution as $Q_f \to 0$. This limit is smooth and results in the unflavored solution corresponding to D2-branes at the tip of a $G_2$-cone, with $\chi=0$. The dilaton then reads 
\begin{equation}
e^{\Phi}\,=\,h^{\frac14}\,=\,\left(c_h+\frac{Q_c}{5\,r^5}\right)^\frac14\,.
\end{equation}
We see that $c_h$ is precisely the constant in the harmonic function of the D2-branes. Setting $c_h=1$ selects asymptotically flat boundary conditions. Here we will instead set $c_h=0$, since this  implements the decoupling limit that yields the equivalence between the gauge theory and the gravity descriptions. 
In conclusion, after fixing the integration constants, the only parameters specifying the solution for massless quarks are the dimensionful 't Hooft coupling $\lambda$ and the dimensionless numbers of D2- and D6-branes $\nc$ and $\nf$, respectively. 

Let us now examine the UV and the IR limits of the solution. The UV regime of the gauge theory corresponds to the region $r\to\infty$, in which the leading-order asymptotic form of the metric and the dilaton are 
\begin{eqnarray}
\dd s_{\rm s}^2&=&\left(\frac{Q_c}{5\,r^5}\right)^{-\frac12}\,\dd x^2_{1,2}+\left(\frac{Q_c}{5\,r^5}\right)^{\frac12}\left(\dd r^2+r^2\,\dd s_6^2\right)\,,\nonumber\\[2mm]
e^{\Phi}&=&\left(\frac{Q_c}{5\,r^5}\right)^{\frac14} \,.
\end{eqnarray}
This is exactly the solution for $\nc$ D2-branes, meaning that the addition of flavor does not modify the UV properties of the theory at leading order. The first corrections in e.g.~the dilaton behave as:
\begin{equation}
e^{\Phi}\,=\,\left(\frac{Q_c}{5\,r^5}\right)^{\frac14}\,
\left[1-\frac{59}{24}\,\frac{Q_f}{r}+\frac{14657}{2688}\,\left(\frac{Q_f}{r}\right)^{2}+\mathcal{O}\left(\frac{Q_f}{r}\right)^{3}\right]\,.
\end{equation}

The IR regime of the gauge theory corresponds to the region $r\to0$, in which the metric and the dilaton take the form
\begin{eqnarray}
\dd s_{\rm s}^2&=&\frac{\rho^2}{L^2}\,\dd x^2_{1,2}+\frac{L^2}{\rho^2}\,\dd \rho^2+\frac94\,L^2\,\dd s_6^2\,,\nonumber\\[2mm]
e^{\Phi}&=&\frac{1}{2\sqrt{3}}\,\left(\frac{Q_c}{Q_f^5}\right)^\frac14\,=\,\frac{1}{g_s}\,\frac{1}{2\sqrt{3}}\,\left(\frac{V_2^5}{V_6}\right)^\frac14\,\left(\frac{\nc}{N_f^5}\right)^\frac14\,,
\label{FP}
\end{eqnarray}
where we have changed  coordinates via $r^3 = Q_f\, \rho^2$ and we have defined
\begin{equation}
L\,=\,\frac{2}{3\sqrt{3}}\,\left(\frac{Q_c}{Q_f}\right)^\frac14\,=\,\frac{4\pi\ell_s}{3\sqrt{3}}\,\left(\frac{V_2}{V_6}\right)^\frac14\,\left(\frac{\nc}{N_f}\right)^\frac14\,.
\end{equation}
Remarkably, the IR geometry is  $AdS_4\times{\rm NK}$ with radius $L$.
This falls in the general class of supersymmetric $AdS_4$ solutions with $SU(3)$-structure internal manifold found in \cite{Lust}. The requirement that type IIA supergravity provides a reliable description, namely the conditions that $L\gg \ell_s$ and $g_s e^\Phi \ll 1$, translate into 
\begin{equation}
N_f\ll \nc\ll N_f^5\,.
\end{equation}
Note that these inequalities require both $N$ and $N_f$ to be large.

There exists a technical difficulty with uplifting to M-theory type IIA solutions in which the Bianchi identity for $F_2$ is violated, but presumably this uplift can be obtained along the lines proposed in \cite{Gaillard:2009kz}. Assuming the usual relation between the radii in ten and eleven dimensions and the dilaton, 
the $AdS$ solution would correspond in M-theory to another $AdS_4$ geometry with radius 
\begin{equation}
L\sim \left(\nc\,N_f\right)^{\frac16} \ell_p\,,
\end{equation}
with $\ell_p$ the eleven-dimensional Planck length. This is always large in the large-$\nc$ limit, thus extending the range of validity of the solution to arbitrary $N_f$ --- see a closely related discussion in \cite{Pelc:1999ms}.

We conclude that  the solution (\ref{massless}) (with $c_\chi=1$ and $c_h=0$)  describes a set of $\mathcal{N}=1$ supersymmetric RG flows (one for each possible internal NK manifold) from a three-dimensional SYM theory in the UV to an interacting fixed point in the IR, driven by the addition to the theory of massless quarks. We show in Appendix \ref{details} that the fixed point is approached along irrelevant directions corresponding to operators of dimensions $\Delta\,=\,6,\,11/3$.

Several observations can be made about the  gauge theories dual to the solutions above. First of all there is the fact that the flow drives the theory to an interacting IR fixed point. The existence of non-trivial conformal theories at the IR of a three-dimensional gauge theory in the presence of a large number of flavors was first observed in \cite{Appelquist}. This was proven in an expansion in $1/N_f$ to all orders. Here we find that this feature is already present in the Veneziano limit, at least for the $\mathcal{N}=1$ cases at hand. To elaborate on this point, we note that the running gauge coupling $g^2(\mu)$ can be computed introducing a D2-brane probe in the flavored background at a fixed radial position $r$, expanding the D2-brane action to quadratic order in the Born-Infeld field strength and matching with the canonically normalized YM term. The result for the dimensionless effective coupling is 
\begin{equation}
g_{\rm eff}^2\equiv\frac{g^2(\mu)N }{\mu}\,=\,\frac{\lambda}{\mu}\,\left(1+\frac{2}{V_2}\,\frac{N_f}{N} \frac{\lambda}{\mu}\right)^{-\frac32}\,.
\end{equation}
To arrive at this equation we have used \eqq{coupling} and the fact that the energy scale in the gauge theory is related to the radial position in the bulk through $2\pi\mu=r/\ell_s^2$. In the UV and in the IR this behaves as 
\be
\mbox{UV:} \qquad g_{\rm eff}^2 \sim \frac{\lambda}{\mu} \sac
\mbox{IR:} \qquad g_{\rm eff}^2 \sim \left(\frac{N}{N_f}\right)^{\frac32}
\sqrt{\frac{\mu}{\lambda}}   \,.
\ee
The UV behavior is that expected of SYM, i.e.~\eqq{geff}. In between these two asymptotic behaviors the coupling attains a unique maximum at 
\begin{equation} 
\mu_{\text{\tiny{CFT}}}\,=\,\frac{\lambda}{V_2}\,\frac{N_f}{\nc}\,,
\label{scale}
\end{equation}
where the subscript is a reminder that the physics below this scale is approximately conformal.
At this scale\footnote{This can be compared with the one-loop result, robust in the large $N_f$ limit (see e.g.~the discussion in \cite{Polchinski:2010hw})
\begin{equation}
\mu\partial_\mu g_{\rm eff}^2\,=\,-g_{\rm eff}^2+\beta_0\,g_{\rm eff}^4
\end{equation}
for some positive $\beta_0$. This vanishes at a non-zero coupling $g_{\rm eff}^2=1/\beta_0$.}
\begin{equation}
g_{\rm eff}^2\,=\,\frac{V_2}{3\sqrt{3}}\,\frac{N}{N_f}\,.
\end{equation}
and the YM $\beta$-function vanishes. This may seem to suggest that the theory reaches  a fixed point at a finite energy scale. However, this is not true because the IR YM interactions are subdominant with respect to the Chern--Simons (CS) interactions generated along the flow, which is consistent with the fact that the effective YM coupling actually goes to zero in the deep IR.  
At the perturbative level, the addition of quarks is known to induce a CS term with level proportional to the number of flavors running in the loop \cite{Niemi:1983rq,Redlich}. At strong coupling, this effect can be seen in the non-vanishing of the Wess--Zumino (WZ) action for a probe D4-brane that fills out the Minkowski directions and is suitably oriented along the internal directions in the flavored background: 
\begin{equation}
S_{\rm WZ}\,\sim\,
\int A\wedge F\wedge\,\int F_2\,\sim\,
N_f \,\int A\wedge F\,,
\end{equation}
where $F$ is the field strength of the gauge field $A$ living on the brane. We see that the presence of an $F_2$ flux induces a CS level proportional to $\nf$.

It has been conjectured that the free energy of a three-dimensional (Euclidean) field theory placed on the three-sphere is a genuine measure of the number of degrees of freedom and verifies appropriate monotonicity theorems \cite{Jafferis:2011zi}.
This free energy thus plays a role analogous to that of the $c$-function for two-dimensional field theories or the $a$-coefficient in the Weyl anomaly in four dimensions. For a theory at a fixed point with a holographic description, the free energy is proportional to the $AdS_4$ radius measured in units of the four-dimensional effective Planck length. For the IR fixed point 
\eqq{FP} the free energy scales with the number of flavors and the rank of the gauge group as
\begin{equation}
F\left(S^3\right)\, \sim\,\frac{L^2\,e^{-2\Phi}}{2\kappa_4^2}\,\sim\,\frac{L^8\,e^{-2\Phi}}{2\kappa^2}\,\sim\,\left(\nc^3\,N_f\right)^{\frac12}\,,
\end{equation} 
where the dilaton factors account for the fact that the metric \eqq{FP} is written  in string frame. We see that the $N$-dependence (the adjoint contribution) matches the peculiar power of CS-matter theories of the ABJM type. 

The consistent picture that emerges from this analysis is that, below the scale 
\eqq{scale}, the CS term dominates over the YM term and the IR dynamics of our models is governed by a CS-matter theory. In the case in which the internal NK manifold is $\mathbb{CP}^3$ this can be made more explicit because of the connection with the solution by Ooguri and Park (OP) \cite{Ooguri}, which is an $\mathcal{N}=1$ deformation of the ABJM \cite{Aharony} theory. Indeed, Conde and Ramallo (CR) \cite{Conde} constructed solutions corresponding to the addition of backreacting flavor to both the ABJM solution and the OP solution.  Regarding $\mathbb{CP}^3$  as an $S^2$ fibration over an $S^4$ base --- see the paragraph below \eqq{NKlist} --- CR showed that the backreaction produces a relative squashing between the fiber and the base, as well as a deformation of the RR form $F_2$ to accommodate the violation of the Bianchi identity. CR  parametrize this squashing and this deformation with two dimensionless quantities $q$ and $\eta$, respectively. Given $\eta$, $q=q(\eta)$ is determined by a second-order algebraic equation, which results in two branches of solutions. Moving along a given branch corresponds to changing the number of flavors. Crucially, the ABJM  and the OP solutions lie on different branches. By explicit construction of the NK structure on $\mathbb{CP}^3$, it can be shown that our IR fixed point corresponds to a solution with $q=2, \eta=-2$ in the language of CR, and this solution happens to be on the same branch as the unflavored OP solution. In fact, the results of CR show that this solution is obtained by adding $\nf=4|k|$ flavors to the OP solution. It is interesting to note that the metric on the 
$\mathbb{CP}^3$ manifold of the unflavored OP solution is not even Einstein, yet under the addition of an appropriate number of flavors the metric becomes not just Einstein but NK.

%%%%%%%%%%%%%%%%%%%%%%%%%%%%%%%%%%%%%%%%%%%%%%%%%%%%%%%%%%
\section{Massive quarks and quasi-conformal dynamics}

In the case of massive quarks the D6-branes are separated from the D2-branes by a finite distance proportional to the quark mass. This translates into the fact that the D6-branes extend from infinity down to a non-zero minimal value of the radial coordinate $r_m$. By Gauss' law this means that the D6-brane charge $Q_f(r)$ vanishes for $r<r_m$. In the gauge theory this corresponds to the statement that in the IR, i.e.~at energies below the quark mass, the quarks decouple from the dymanics. In contrast, in the UV the quarks can be treated as effectively massless, so $Q_f(r)$ must approach the value that it would have had in the case of exactly massless quarks. It is therefore convenient to write 
\begin{equation}
Q_f(r)\,=\,Q_f\,p(r)\,,
\end{equation}
where $Q_f$ is now a constant and  $p(r)$ is a dimensionless function that vanishes for $r<r_m$ and that approaches 1 as $r\to \infty$. In between these two limits $p(r)$ is monotonically increasing. 

It is important to note that $p(r)$ is not determined dynamically: any  function with the  properties mentioned above is an admissible choice. On the gauge theory side, this freedom corresponds to the freedom of adding quarks of many different masses to the theory.  The supergravity equations can be solved for any $p(r)$ in terms of an integral for $\chi$ and a double integral for $h$ as
\begin{eqnarray}
\label{general}
e^{-2\chi(r)}&=&c_\chi+2\,Q_f\,\int_r^\infty\,p(z)\,\frac{\dd z}{z^2}\,,\nonumber\\[4mm]
h(r)&=&e^{-3\chi}\,\left(c_h+Q_c\,\int_r^\infty\,\,e^{\chi(y)}\,\,\frac{\dd y}{y^6}\right)\,.
\end{eqnarray}
Taking the massless limit $p\to1$, one can see that the integration constants have the same meaning as before, so we can set $c_{\chi}=1$ and $c_h=0$. 

There is  a family of functions $p(r)$, labelled by $r_m$, such that each member of the family describes a theory in which all quarks have exactly the same mass proportional to $r_m$. In order to find this family some details about the internal manifold are needed. This means that to proceed further we cannot treat all the NK geometries simultaneously. For illustrative purposes, we will therefore focus on the six-sphere henceforth. 

The special class of functions corresponding to quarks of equal masses is determined by the following consistency condition. By assumption, the full supergravity solution includes the backreaction of a large number of flavor branes smeared over the internal manifold by the action of a symmetry group --- $G_2$ in the case of D6-branes on the $S^6$. Each of these branes is embedded non-trivially on the internal manifold, but all these embeddings are related by a symmetry. This means that the action for the entire set of branes should equal $\nf$ times the action of a single one. The mathematical expression of this statement is 
\begin{equation}
T_{\rm D6} \int_{10}\,e^{-\Phi}\,\mathcal{K}\wedge\Xi\,=\,
N_f \, T_{\rm D6} \int_7\,e^{-\Phi}\,\mathcal{P}\left[\mathcal{K}\right]_{\rm D6}\,,
\label{state}
\end{equation}
where $\mathcal{P}$ denotes the pullback to the brane. The action on the left-hand side is integrated over the entire spacetime and depends on the D6-brane embedding indirectly only through $p(r)$. The action on the right-hand side is only integrated over the seven dimensional submanifold occupied by a fiducial D6-brane and depends explicitly on the embedding of the brane in spacetime.  Thus this equation will relate $p(r)$ to the embedding of the brane and hence to the quark mass.  Note that we should in principle consider the full action instead of just the Dirac-Born-Infeld (DBI) part, but this is not necessary in this case because the analysis for the Wess-Zumino (WZ) part of the action follows from the DBI analysis by supersymmetry.  

Using our calibration and smearing forms, the left-hand side of \eqq{state} 
can be easily computed with the result
\begin{equation}
\label{1}
T_{\rm D6}\,\int_{10}\,e^{-\Phi}\,\mathcal{K}\wedge\Xi\,=\,\frac{12\,V_6\,Q_f}{2\kappa^2}\,\int\,e^\chi\,r^3\,\left(p+\frac{r}{4}\,p'\right)\,\dd^3 x\dd r\,,
\end{equation}
where we have explicitly performed the integration over the internal manifold. 

In order to preserve supersymmetry, the three-cycle wrapped by a D6-brane  inside a NK manifold must be calibrated by  ${\rm Re}\,\Omega$ \cite{Gutowski, Koerber}. In the case of the six-sphere an equatorial ${S}^3\subset{S}^6$ provides an example of such a cycle, so it is convenient to  write the metric on the six-sphere as
\begin{equation}
\dd \Omega^2_6\,=\,\dd \theta^2+\sin^2{\theta}\,\dd\Omega_3^2+\cos^2\theta\,\dd\Omega_2^2\,,
\end{equation}
where $\dd \Omega_n^2$ denotes the metric of the unit-radius $n$-sphere. The D6-brane embedding can then be specified as $\theta=\theta(r)$. To compute the pullback of the calibration form one needs to use the following two results, which are easily obtained  using the explicit realization of the NK structure presented in Appendix \ref{NKS6}:
\begin{eqnarray}
\mathcal{P}\left[{\dd^3x\wedge\dd r\wedge\rm Re}\,\Omega\right]_{\rm D6}&=&\sin^4\theta\,\,\dd^3x\wedge\dd r\wedge\epsilon_{(3)}\,,\nonumber\\[2mm]
\mathcal{P}\left[\dd^3x\wedge J\wedge J\right]_{\rm D6}&=&2\cos\theta\,\sin^3\theta\,\,\theta'\,\,\dd^3x\wedge\dd r\wedge\epsilon_{(3)}\,,
\end{eqnarray}
with prime denoting the radial derivative and $\epsilon_{(3)}$ being the volume form of the equatorial $S^3$. The right-hand side of \eqq{state} then takes the form
\begin{equation}\label{smearedDBI}
T_{\rm D6}\,\int\,e^{-\Phi}\,\mathcal{P}\left[\mathcal{K}\right]_{\rm D6}\,=\,T_{\rm D6}\,V_3\,\int\,e^{\chi}\,r^3\,\left(\sin^4\theta+\frac{r}{4}\,\left(\sin^4\theta\right)'\right)\,\dd^3x\dd r\,,
\end{equation}
where $V_3$ is the volume of the S$^3$ wrapped by the brane. Comparing \eqq{1} and \eqq{smearedDBI} and using the quantization condition (\ref{quantcond}), together with the fact that $12 V_6=V_2 V_3$, we arrive at the conclusion that 
\begin{equation}\label{correctp}
p(r) = \sin^4\theta(r) \,.
\end{equation}
As anticipated, this equation relates the D6-brane embedding $\theta(r)$ with the supergravity charge density function $p(r)$. The quark mass enters as a boundary condition on the D6-brane embedding, so our next task is to determine the equation of motion and the boundary conditions for the latter. 

The brane embedding is determined by the consistency condition that it solves the equation of motion of a D6-brane probe in the background generated by the D6-branes themselves. The DBI part of the action for the probe takes the form
\begin{equation}\label{DBID6}
S_{\rm DBI}\,=\,-T_{\rm D6}\,\int\,e^{-\Phi}\,\sqrt{-g_{\rm D6}}\,\dd^7\xi\,=\,-T_{\rm D6}\,V_3\,\int\,e^{\chi}\,r^3\,\sin^3\theta\left(1+r^2\,\theta'^2\right)^{\frac12}\,\dd^3x\dd r\,.
\end{equation}
The WZ part of the action depends on the background RR 
seven-form. This is determined by the condition that $\dd C_7\,=\,F_8\,=\,*F_2$, which is solved by\footnote{Gauge invariance implies that $C_7$ is only defined up to exact terms. In view of the calibration condition (\ref{calcond}), we could have also chosen $C_7=e^{-\Phi}\,\mathcal{K}$. The choice we adopted differs from this one by an exact piece.}
\begin{equation}
C_7\,=\,-Q_f\,e^{3\chi}\,\frac{r^2}{4}\,\dd^3x\wedge\dd r\wedge{\rm Re}\,\Omega\,.
\end{equation}
The WZ part of the action is now easily calculated with the result 
\begin{equation}\label{WZD6}
S_{\rm WZ}\,=\,T_{\rm D6}\,\int\,\mathcal{P}\left[C_7\right]_{\rm D6}\,=\,-T_{\rm D6}\,V_3\,\int\,Q_f\,e^{3\chi}\,\frac{r^2}{4}\,\sin^4\theta\,\dd^3x\dd r\,.
\end{equation}
In principle, we should now solve the second-order equations of motion that follow from varying $S_{\rm DBI} + S_{\rm WZ}$ with respect to $\theta(r)$. However, in this case we can omit this step because we know that the D6-brane embedding is supersymmetric, which means that the brane embedding will be determined by a first-order BPS equation. Moreover, because of the no-force condition between different D6-branes, we expect this equation to be the same as for a D6-brane probe in an unflavoured  background sourced by D2-branes alone. This equation is known to be 
\begin{equation}
\theta'\,=\,\frac{\cot\theta}{r}\,,
\label{holds}
\end{equation}
and it is easy to verify that the second-order equations of motion are automatically satisfied provided \eqq{holds} holds. The solution for the brane embedding is thus 
\begin{equation}
\cos\theta\,=\,\frac{r_m}{r}\,,
\end{equation}
where $r_m$ is an integration constant that determines both the asymptotic behavior of the brane embedding and the lowest value of the radial coordinate attained by the brane. From the asymptotic behavior we read off that the relation between $r_m$ and the bare quark mass that enters the UV gauge theory Lagrangian is \cite{Karch:2002sh,Kruczenski:2003be} 
\be
m_q = \frac{r_m}{2\pi\ell_s^2} \,.
\ee

Having solved for the brane embedding we can now use  (\ref{correctp}) to obtain  the charge distribution
\begin{equation}
p(r)\,=\,\left[1-\left(\frac{r_m}{r}\right)^2\right]^2\,\Theta\left(r-r_m\right)\,,
\end{equation}
where $\Theta$ is the Heaviside theta function. Substituting into \eqq{general} we obtain the supergravity solution for massive quarks:
\begin{eqnarray}\label{massivesol}
e^{-2\chi}&=&\left\{\begin{array}{ll}
		1+\frac{16}{15}\,\frac{Q_f}{r_m}  & \qquad{\rm if } \,\,\, r<r_m \\[4mm]
		1+\frac{2\,Q_f}{r}\,\left(1-\frac23\left(\frac{r_m}{r}\right)^2+\frac15\left(\frac{r_m}{r}\right)^4\right) &\qquad{\rm if } \,\,\, r\ge r_m
	\end{array}
\right.\nonumber\\[4mm]
h&=&\left\{\begin{array}{ll}
		\left(1+\frac{16}{15}\,\frac{Q_f}{r_m}\right)\,\frac{Q_c}{5\,r^5}  & \qquad{\rm if } \,\,\, r<r_m \\[4mm]
		e^{-3\chi}\,Q_c\,\int_r^\infty\,e^{\chi(y)}\,\,\frac{\dd y}{y^6} &\qquad{\rm if } \,\,\, r\ge r_m\,.
	\end{array}
\right.
\end{eqnarray}
The constants of integration have been adjusted so that in the UV the D2-brane solution is recovered, as for the massless case. In the IR, below the scale set by $r_m$, the D2-brane solution is also recovered, but with a different normalization that translates into a finite renormalization of the coupling: 
\be
\lambda_{\text{\tiny{IR}}} = \left(1+\frac{16}{15}\,\frac{Q_f}{r_m}\right)^{-1}\,\lambda_{\text{\tiny{UV}}}\,.
\ee
This difference in normalizations  is needed to ensure the continuity of the solution at $r_m$.\footnote{In fact the solution at $r_m$ is not just continuous  but ${\cal C}^2$.} The quark mass is similarly renormalized. This can be seen by computing the effective quark mass at the decoupling scale, which can be read off from the action  of a string stretching between the lowest point on a D6-brane and the D2-branes: 
\begin{equation}\label{quarkmass}
m_q^{\text{\tiny{IR}}}\,=\,\frac{1}{2\pi\ell_s^2}\,\int_0^{r_m}\,\sqrt{-g_{tt}g_{rr}}\,\dd r\,=\,\frac{1}{2\pi\ell_s^2}\,\int_0^{r_m}\,e^{\chi}\,\dd r\,=\,\frac{r_m}{2\pi\ell_s^2}\,\left(1+\frac{16}{15}\,\frac{Q_f}{r_m}\right)^{-\frac12}\,.
\end{equation}
Note that $m_q^{\text{\tiny{IR}}} \leq m_q$.

\begin{figure}[h!!!]
\begin{center}
\includegraphics[width=0.65\textwidth]{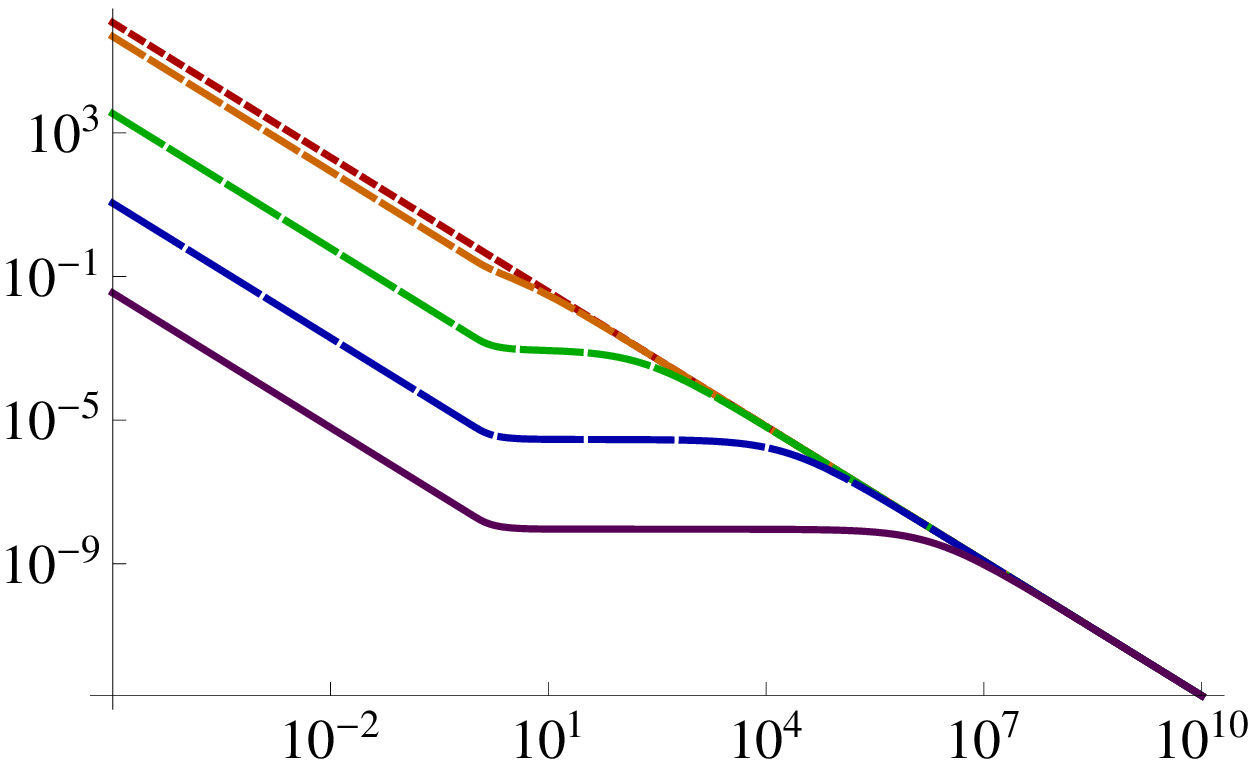}
\put(-290,170){$e^\Phi$}
\put(10,4){$r/r_m$}
\caption{\small \label{Massivedilaton}
Dilaton as a function of the energy scale for the  solutions with massive quarks. The values of $r_m$ and $g_s$ on the gravity side, or equivalently of $m_q$ and $\lambda$ on the gauge theory side, are the same for all curves. Instead, the ratio $N_f/\nc$ increases from top to bottom. Specifically, we take $Q_c=r_m^5$ and $Q_f=r_m 10^n$ with $n=\{-2,0,2,4,6\}$ from top to bottom, which translates into $\nf/N = 10^n \, m_q / \lambda$.}
\end{center}
\end{figure}

The solution exhibits two  qualitatively different regimes depending on the value of the ratio $Q_f/r_m$, which in terms of gauge theory parameters may be written as
\begin{equation}
\frac{Q_f}{r_m}\,\sim\,\frac{\lambda}{m_q}\,\frac{N_f}{\nc} 
\sim \frac{\mu_{\text{\tiny{CFT}}}}{m_q} \,,
\end{equation}
with $\mu_{\text{\tiny{CFT}}}$ the scale introduced in \eqq{scale}. If $m_q \ll \mu_{\text{\tiny{CFT}}}$ then the theory first reaches the region at $\mu \sim \mu_{\text{\tiny{CFT}}}$ in which the physics is described by a conformal CS-matter theory, as described in Section \ref{masslessquarks}, and only at a much lower scale $m_q$ it `realizes' that the quark mass is non-zero. Thus in this case  the theory exhibits `walking' at scales $m_q \ll \mu \ll \mu_{\text{\tiny{CFT}}}$, i.e.~the physics is approximately conformal in this window. In contrast, if $m_q \gtrsim \mu_{\text{\tiny{CFT}}}$, then the quarks decouple from the dynamics before their presence can drive the theory to an approximately conformal phase and the walking region disappears. These two regimes are clearly seen in the behavior of the dilaton displayed in Fig.~\ref{Massivedilaton}, obtained by numerically integrating (\ref{massivesol}).
The outcome only depends on the ratio between scales, which are clearly visible in the plot. The lower one is approximately $m_q$, that we keep fixed. The 't Hooft coupling $\lambda$ is also common to all curves in order to have the same UV asymptotics. The different curves are only distinguished by the value of $\nf/N$, which increases from top to bottom. In the bottom curves a quasi-conformal region in which the dilaton is approximately constant is clearly visible.  As we decrease $N_f/\nc$ and move to the  top curves, the size of the walking region decreases until it disappears all together.

%%%%%%%%%%%%%%%%%%%%%%%%%%%%%%%%%%%%%%%%%%%%%%%%%%%%%%%%%%%%%%
\section{Outlook}\label{conclusions}

We have constructed analytic solutions dual to three dimensional SYM theories with dynamical flavors of arbitrary mass. The simplicity of the solutions, in particular the fact that  they can be found by solving ordinary differential equations, opens the door to several interesting extensions including the addition of temperature and/or a quark density. In the first case we expect that, as the ratio $m_q/T$ is varied,  the solutions will exhibit the backreacted version of the phase transitions between Minkowski and black hole embeddings uncovered in  \cite{Babington:2003vm,Mateos:2006nu,Mateos:2007vn}. In these second case we can make contact  \cite{prep} with the results of \cite{Faedo:2014ana} in the limit in which the quarks mass becomes sufficiently large.

%%%%%%%%%%%%%%%%%%%%%%%%%%%%%%%%%%%%%%%%%%%%%%%%%%%%%%%%%%%%%%
\section*{Acknowledgments}

\noindent We are grateful to Davide Cassani, Eduardo Conde, Carlos N\'u\~nez and Alfonso Ramallo. We are supported by grants 2014-SGR-1474, MEC MEC FPA2013-46570-C2-1-P, MEC FPA2013-46570-C2-2-P, CPAN CSD2007-00042 Consolider- Ingenio 2010, and ERC Starting Grant HoloLHC-306605. JT is also supported by the Juan de la Cierva program of the Spanish Ministry of Economy.

%%%%%%%%%%%%%%%%%%%%%%%%%%%%%%%%%%%%%%%%%%%%%%%%%%%%%%%%%%%%%%
\appendix

%%%%%%%%%%%%%%%%%%%%%%%%%%%%%%%%%%%%%%%%%%%%%%%%%%%%%%%%%%%%%%
\section{Conventions}\label{details}

In this appendix we provide the details needed to reproduce the solutions in the main text together with some additional material, like the reduction of the system to a four-dimensional action. 

%%%%%%%%%%%%%%%%%%%%%%%%%%%%%%%%%%%%%%%%%%%%%%%%%%%%%%%%%%%%%%
\subsection{Ten-dimensional action and equations}

We work with type IIA supergravity in string frame, enlarged with adequate sources. The total action is:
\begin{equation}
S=S_{\rm IIA}+S_{\rm sources}=S_{\rm NS}+S_{\rm RR}+S_{\rm sources}\,.
\end{equation}
The first part is the action of type IIA supergravity. The Neveu--Schwarz sector is
\begin{equation}
S_{\rm NS}\,=\,\frac{1}{2\kappa^2}\,\int\,e^{-2\Phi}\left(R*1+4\dd\Phi\wedge*\dd\Phi-\frac12H\wedge*H\right)\,,
\end{equation}
with $H=\dd B$ solving the Bianchi $\dd H=0$. The RR piece, containing the kinetic terms for the two- and four-forms plus a topological interaction (whose explicit form will not be needed here), reads 
\begin{equation}
S_{\rm RR}\,=\,\frac{1}{2\kappa^2}\,\int\,\left(-\frac12F_2\wedge*F_2-\frac12F_4\wedge*F_4\right)+S_{\rm top}\,.
\end{equation}
In the absence of sources, the field strengths are
\begin{equation}
F_2\,=\,\dd C_1\,,\qquad\qquad \qquad F_4\,=\,\dd C_3-H\wedge C_1\,,
\end{equation}
solving the Bianchi identities
\begin{equation}
 \dd F_2\,=\,0\,,\qquad\qquad\qquad \dd F_4\,=\,H\wedge F_2\,.
\end{equation}
On the other hand, the action for the sources is the sum of the smeared DBI and WZ terms\footnote{Here and in all subsequent equations we assume that the gauge invariant combination ${\cal F}=2\pi\alpha' dA+B$ vanishes, where $A$ is the Born-Infeld field on the branes.}
\begin{equation}
S_{\rm sources}\,=\,-T_{\rm D6}\,\int\,\left(e^{-\Phi}\,\mathcal{K}-C_7\right)\wedge\Xi\,,
\end{equation}
where $\mathcal{K}$ is the calibration form, given essentially by the induced metric on the brane, and $C_7$ is the Hodge dual of $C_1$, with the convention
\begin{equation}
\begin{array}{rclcrcl}
F_2&=&-*F_8\quad\quad&\Rightarrow&\quad\quad F_8&=&*F_2\\[3mm]
F_4&=&*F_6\quad\quad&\Rightarrow&\quad\quad F_6&=&-*F_4\,.
\end{array}
\end{equation}
This new coupling between the D6-branes and $C_7$ is translated into a modification of the Bianchi for $F_2$ (which coincides with the equation of motion for $F_8$) that now reads
\begin{equation}
\dd F_2\,=\,-2\kappa^2T_{\rm D6}\,\Xi\,.
\end{equation}
The equations of motion for the lower forms are unmodified by the sources
\begin{eqnarray}
\dd\left(e^{-2\Phi}*H\right)-F_2\wedge*F_4-\frac12F_4\wedge F_4&=&0\nonumber\\[4mm]
\dd *F_2+H\wedge*F_4&=&0\\[4mm]
\dd*F_4+H\wedge F_4&=&0\nonumber\,.
\end{eqnarray}
In contrast, the equation of motion for the dilaton receives  an additional contribution from the DBI action
\begin{equation}
R*1+4\dd*\dd\Phi-4\dd\Phi\wedge*\dd\Phi-\frac12H\wedge*H-\kappa^2\,T_{\rm D6}\,e^\Phi\,\mathcal{K}\wedge\Xi\,=\,0\,.
\end{equation}
Einstein's equations are also modified to reflect the presence of the explicit sources
\begin{equation}
R_{MN}+2\nabla_M\nabla_N\Phi-\frac14H_{MRS}H_N{}^{RS}\,=\,T_{MN}^{\rm IIA}+T_{MN}^{\rm sources}\,,
\end{equation}
where $T_{MN}^{\rm IIA}$ is the usual energy-momentum tensor of the RR sector\footnote{For arbitrary $n$-forms $\omega_n$ and $\xi_n$, we have defined the contraction symbol as $\omega_n\lrcorner\xi_n=\frac{1}{n!}\omega_{M_1\dots M_n}\xi^{M_1\dots M_n}$.}
\begin{equation}
T_{MN}^{\rm IIA}\,=\,e^{2\Phi}\,\left[\frac12(F_2)_{MR}(F_2)_N{}^R+\frac{1}{12}(F_4)_{MR_1R_2R_3}(F_4)_N{}^{R_1R_2R_3}-\frac14g_{MN}\left(F_2\lrcorner F_2+F_4\lrcorner F_4\right)\right]\,.
\end{equation}
The contribution from the sources is
\begin{equation}
T_{MN}^{\rm sources}\,=\,\frac{\kappa^2\,T_{\rm D6}}{2}\,e^\Phi\,\left[g_{MN}\,\Xi\lrcorner\left(*\mathcal{K}\right)-\Xi_{MRS}\left(*\mathcal{K}\right)_N{}^{RS}\right]\,.
\end{equation}
In deriving the last equation we used the equation of motion for the dilaton, as well as the facts that  $\mathcal{K}\wedge\Xi=-\left[\Xi\lrcorner\left(*\mathcal{K}\right)\right]*1$ depends on the metric only through the vielbeins and that 
\be
\frac{\delta}{\delta g^{MN}}=\frac12e^A_N\frac{\delta}{\delta e^{AM}} \,. 
\ee
The WZ term in the action does not contribute to  this equation, since it is topological, its only effect being the modification of the Bianchi identity for the two-form.

%%%%%%%%%%%%%%%%%%%%%%%%%%%%%%%%%%%%%%%%%%%%%%%%%%%%%
\subsection{BPS equations and reduction to four dimensions}

The conditions for having $\mathcal{N}=1$ solutions preserving three-dimensional Poincar\'e invariance with a NK internal geometry in type IIA were derived in \cite{Haack}. We can capitalize on their results to write the BPS equations for our system simply by allowing for a suitable violation of the Bianchi identity for the 2-form. In order to make contact with their notation, here we use the following ansatz for the NS sector\footnote{With the exception that we have changed to a more convenient radial variable $\rho\, \dd r_{\rm there}=\tau\, \dd z_{\rm here}$.} 
\begin{eqnarray}
\dd s_{\rm s}^2&=&\frac{e^{2A}}{\tau^2}\dd x^2_{1,2}+\rho\,\left(\dd z^2+\dd s_6^2\right)\,,\nonumber\\[4mm]
e^{\Phi}&=&\frac{\rho^{\frac32}}{\tau}\,,\qquad\qquad\qquad H\,=\,0\,,
\end{eqnarray}
where $\dd s_6^2$ is the metric of the internal NK manifold, normalized to $R_{ab}=5\,\delta_{ab}$. The RR-forms are taken as in Eq.~(\ref{RRansatz}). Then, the set of BPS equations that can be gathered from \cite{Haack} by including the effect of the flavor is
\begin{eqnarray}\label{BPSset}
\frac{\rho'}{\rho}&=&2-\frac{Q_c}{2}\,\frac{1}{\rho\,\tau}+\frac{Q_f}{2}\,\frac{\rho}{\tau}\nonumber\\[4mm]
\frac{\tau'}{\tau}&=&3-\frac{Q_c}{2}\,\frac{1}{\rho\,\tau}-\frac{3\,Q_f}{2}\,\frac{\rho}{\tau}\\[4mm]
A'&=&3-\frac{Q_c}{4}\,\frac{1}{\rho\,\tau}-\frac{3\,Q_f}{4}\,\frac{\rho}{\tau}\nonumber\,.
\end{eqnarray}
As usual, the equation for the warp factor can be integrated in terms of the rest of the functions. In this case it can be written as
\begin{equation}
A\,=\,A_0+\frac32\,z+\frac12\,\log\tau\,,
\end{equation}
where $A_0$ is a content-free integration constant that can be absorbed into a rescaling of the Minkowski coordinates. Thus, the information of the system is contained in just two functions, which justifies the form of our ansatz in the bulk of the paper. The dictionary with the variables used there is
\begin{equation}
\rho\,=\,r^2\,h^\frac12\,e^{2\chi}\,,\qquad\qquad\qquad\tau\,=\,r^3\,h^\frac12\,,\qquad\qquad\qquad e^z\,=\,r\,.
\end{equation}

It is also convenient to have a reduced action from which the BPS equations follow. The reduction of (massive) type IIA supergravity on an arbitrary NK manifold was performed in \cite{Kashani}. The resulting theory is an $\mathcal{N}=2$ gauged supergravity in four dimensions. From the full collection of modes respecting the NK structure, our solutions excite a very restricted subset comprising the dilaton and the breathing mode, that is, the volume modulus of the internal space. On top of that we have to allow for the deformation coming from the flavor branes, which was not considered in \cite{Kashani}. The reduced four-dimensional action that captures the dynamics of our brane intersection reads\footnote{The unusual term $Q_f'$, proportional to the radial derivative of the flavor charge distribution, is needed in the massive case. Varying with respect to $Q_f$ gives the calibration condition. Here and in the following we will use a four-dimensional radial coordinate, related to the ten-dimensional one by $\rho^{\frac12}\,\tau\,\dd z_{10}=\dd z_{4}$.}
\begin{eqnarray}\label{4daction}
S_4&=&\frac{1}{2\kappa^2}\,\int\,\left(R*1-\frac{3}{2\rho^2}\,\dd \rho\wedge*\dd \rho-\frac{2}{\tau^2}\,\dd\tau\wedge*\dd\tau-V*1-\frac{3\,\rho^{\frac12}}{\tau^2}\,Q_f'\,*1\right)\nonumber\\[4mm]
&=&\frac{1}{2\kappa^2}\,\int\,\left(R*1-G_{ij}\,\dd\phi^i\wedge*\dd\phi^j-V*1-\frac{3\,\rho^{\frac12}}{\tau^2}\,Q_f'\,*1\right)\,,
\end{eqnarray}
where the potential is
\begin{equation}
V\,=\,\frac{Q_c^2}{2}\,\frac{1}{\rho^3\tau^4}+\frac{3\,Q_f^2}{2}\,\frac{\rho}{\tau^4}+Q_f\,\frac{12}{\tau^3}-\frac{30}{\rho\tau^2}\,.
\end{equation}
The first piece, quadratic in $Q_c$, descends from the kinetic term of the four-form in ten dimensions. The last one is due to the curvature of the internal manifold. This two terms will appear in the reduction on any arbitrary Einstein manifold, not necessarily NK. The additional two terms containing $Q_f$ are exclusive of NK geometries in the presence of flavor. The one quadratic in the number of flavors is inherited from the kinetic term of the two-form, while the linear term is due to the DBI action. Notice that, at least when $Q_f'=0$, this is a consistent truncation, in the sense that any solution to the action (\ref{4daction}) can be uplifted to a type IIA solution in the presence of sources as described in the previous section.

The potential can be derived from the superpotential
\begin{equation}
W\,=\,\frac{\rho\,\left(12\,\tau-3\,Q_f\,\rho\right)-Q_c}{4\,\rho^{\frac32}\,\tau^2}
\end{equation}
by means of the standard relation
\begin{equation}
V\,=\,\left(d-2\right)\left[\left(d-2\right)\,G^{ij}\partial_iW\partial_jW-\left(d-1\right)\,W^2\right]\,.
\end{equation}
Here $d$ is the dimension of the bulk, four in our case. Using the domain wall ansatz
\begin{equation}
\dd s_d^2\,=\,e^{2A}\dd x_{1,d-2}^2+\dd z^2
\end{equation}
all the equations of motion are solved by the first order system
\begin{equation}
A'\,=\,\pm W\qquad\qquad\qquad\qquad\phi^{i\,'}\,=\,\mp\left(d-2\right)\,G^{ij}\partial_jW\,.
\end{equation}
The superpotential has an extremum at 
\begin{equation}
\rho\,=\,\frac13\,\left(\frac{Q_c}{Q_f}\right)^\frac12\,,\qquad\qquad\qquad\qquad\tau\,=\,\frac23\,\left(Q_c\,Q_f\right)^\frac12\,,
\end{equation}
which is of course an extremum of the potential as well. This uplifts to a supersymmetric $AdS_4$ geometry, in the class found in \cite{Lust}, that governs the IR of the more general solutions to the full set (\ref{BPSset}). From the derivatives of the potential we can compute the masses of the scalars around the $AdS$ point. They turn out to be 
\begin{equation}
m^2\,L^2\,=\,18,\,\frac{22}{9}\,,
\end{equation}
corresponding to irrelevant operators of dimensions
\begin{equation}
\Delta\,=\,6,\,\frac{11}{3}\,.
\end{equation}

%%%%%%%%%%%%%%%%%%%%%%%%%%%%%%%%%%%%%%%%%%%%%%%%%%%%%%%%%%%%%%
\section{Nearly K\"ahler  structure of $S^6$}\label{NKS6}
%%%%%%%%%%%%%%%%%%%%%%%%%%%%%%%%%%%%%%%%%%%%%%%%%%%%%%%%%%%%%%

In this appendix we provide the details for constructing a NK structure on the six-sphere, needed to obtain the massive quarks solution. Contrary to other NK manifolds, there exist infinitely many such structures compatible with the round metric on the sphere \cite{Butruille}. In \cite{Koerber} a constructive algorithm is given by considering the cone over $S^6$, that is $\mathbb{R}^7$, as the space of octonions. We will simply state the final result in a convenient set of coordinates. 

Let us write the metric of the six sphere as
\begin{equation}
\dd \Omega^2_6\,=\,\dd \theta^2+\sin^2{\theta}\,\dd\Omega_3^2+\cos^2\theta\,\dd\Omega_2^2\,,
\end{equation}
where $\dd\Omega_2^2$ and $\dd\Omega_3^2$ are respectively the metrics of a two and three sphere
\begin{equation}
\dd\Omega_2^2\,=\,\dd \alpha_1^2+\sin^2{\alpha_1}\dd \alpha_2^2\,,\qquad\qquad \dd\Omega_3^2\,=\,\dd \beta_1^2+\sin^2{\beta_1}\left(\dd \beta_2^2+\sin^2{\beta_2}\dd \beta_3^2\right)\,.
\end{equation}
Take the following functions of the angles:
\begin{equation}
\begin{array}{rclcrcl}
u^1&=&\cos\theta\,\cos\alpha_1\,,&\qquad\qquad&u^2&=&\cos\theta\,\sin\alpha_1\,\cos\alpha_2\,,\\
u^3&=&\cos\theta\,\sin\alpha_1\,\sin\alpha_2\,,&\qquad\qquad&u^4&=&\sin\theta\,\cos\beta_1\,,\\
u^5&=&\sin\theta\,\sin\beta_1\cos\beta_2\,,&\qquad\qquad&u^6&=&\sin\theta\,\sin\beta_1\,\sin\beta_2\,\cos\beta_3\,,\\
u^7&=&\sin\theta\,\sin\beta_1\,\sin\beta_2\,\sin\beta_3\,,&&&&
\end{array}
\end{equation}
dictated by the embedding of the unit-radius sphere into flat space, $\sum_n u^n\,u^n=1$. In terms of the one-forms $v^n\equiv\dd u^n$, the metric can be written as $\sum_n v^n\,v^n=\dd \Omega^2_6$, while the almost complex structure reads
\begin{eqnarray}
J&=&u^1\left(v^{23}+v^{47}+v^{56}\right)+u^2\left(v^{45}+v^{67}-v^{13}\right)+u^3\left(v^{12}+v^{57}-v^{46}\right)\nonumber\\
&+&u^4\left(v^{36}-v^{17}-v^{25}\right)+u^5\left(v^{24}-v^{16}-v^{37}\right)+u^6\left(v^{15}-v^{27}-v^{34}\right)\nonumber\\
&+&u^7\left(v^{14}+v^{26}+v^{35}\right)\,.
\end{eqnarray}
On the other hand the complex three-form has real part
\begin{eqnarray}
{\rm Re}\,\Omega&=&u^1\left(v^{257}-v^{246}-v^{345}-v^{367}\right)+u^2\left(v^{146}+v^{356}+v^{347}-v^{157}\right)\nonumber\\
&+&u^3\left(v^{145}+v^{167}-v^{247}-v^{256}\right)+u^4\left(v^{237}+v^{567}-v^{126}-v^{135}\right)\nonumber\\
&+&u^5\left(v^{127}+v^{236}+v^{134}-v^{467}\right)+u^6\left(v^{124}+v^{457}-v^{137}-v^{235}\right)\nonumber\\
&+&u^7\left(v^{136}-v^{125}-v^{234}-v^{356}\right)\,,
\end{eqnarray}
as well as imaginary part
\begin{eqnarray}
{\rm Im}\,\Omega&=&v^{123}+v^{156}+v^{147}+v^{245}+v^{267}-v^{346}+v^{357}\,.
\end{eqnarray}
As mentioned in the main text, a BPS D6-brane has to wrap a three-dimensional submanifold $\Sigma$ calibrated by ${\rm Re}\,\Omega$. This implies ${\rm Im}\,\Omega\,\vline_\Sigma=0$, as shown in \cite{Koerber}. An example is the equatorial $S^3\subset S^6$ given in our coordinates by $\theta=\pi/2$.


\begin{thebibliography}{99}

  \bibitem{Itzhaki}
  N.~Itzhaki, J.~M.~Maldacena, J.~Sonnenschein and S.~Yankielowicz,
  ``Supergravity and the large N limit of theories with sixteen supercharges,''
  Phys.\ Rev.\ D {\bf 58} (1998) 046004
  [hep-th/9802042].
  %%CITATION = HEP-TH/9802042;%%
  
      \bibitem{Acharya2}
  B.~S.~Acharya, J.~M.~Figueroa-O'Farrill, C.~M.~Hull and B.~J.~Spence,
  ``Branes at conical singularities and holography,''
  Adv.\ Theor.\ Math.\ Phys.\  {\bf 2} (1999) 1249
  [hep-th/9808014].
  %%CITATION = HEP-TH/9808014;%%
  
    \bibitem{Karch}
  A.~Karch and E.~Katz,
  ``Adding flavor to AdS / CFT,''
  JHEP {\bf 0206} (2002) 043
  [hep-th/0205236].
  %%CITATION = HEP-TH/0205236;%%
  
  \bibitem{Pelc:1999ms} 
  O.~Pelc and R.~Siebelink,
  ``The D2 - D6 system and a fibered AdS geometry,''
  Nucl.\ Phys.\ B {\bf 558}, 127 (1999)
  [hep-th/9902045].
  
  \bibitem{Cherkis}
  S.~A.~Cherkis and A.~Hashimoto,
  ``Supergravity solution of intersecting branes and AdS/CFT with flavor,''
  JHEP {\bf 0211} (2002) 036
  [hep-th/0210105].
  %%CITATION = HEP-TH/0210105;%%
  
  \bibitem{GomezReino:2004pw} 
  M.~Gomez-Reino, S.~G.~Naculich and H.~Schnitzer,
  ``Thermodynamics of the localized D2-D6 system,''
  Nucl.\ Phys.\ B {\bf 713}, 263 (2005)
  [hep-th/0412015].
  
  \bibitem{Erdmenger:2004dk} 
  J.~Erdmenger and I.~Kirsch,
  ``Mesons in gauge / gravity dual with large number of fundamental fields,''
  JHEP {\bf 0412}, 025 (2004)
  [hep-th/0408113].
  
  
  
  \bibitem{Bigazzi}
  F.~Bigazzi, R.~Casero, A.~L.~Cotrone, E.~Kiritsis and A.~Paredes,
  ``Non-critical holography and four-dimensional CFT's with fundamentals,''
  JHEP {\bf 0510} (2005) 012
  [hep-th/0505140].
  %%CITATION = HEP-TH/0505140;%%
  
    \bibitem{Nunez:2010sf} 
  C.~Nunez, A.~Paredes and A.~V.~Ramallo,
  ``Unquenched Flavor in the Gauge/Gravity Correspondence,''
  Adv.\ High Energy Phys.\  {\bf 2010}, 196714 (2010)
  [arXiv:1002.1088 [hep-th]].
  
  
  \bibitem{Conde}
  E.~Conde and A.~V.~Ramallo,
  ``On the gravity dual of Chern-Simons-matter theories with unquenched flavor,''
  JHEP {\bf 1107} (2011) 099
  [arXiv:1105.6045 [hep-th]].
  %%CITATION = ARXIV:1105.6045;%%
  
  \bibitem{Ooguri}
  H.~Ooguri and C.~S.~Park,
  ``Superconformal Chern-Simons Theories and the Squashed Seven Sphere,''
  JHEP {\bf 0811} (2008) 082
  [arXiv:0808.0500 [hep-th]].
  %%CITATION = ARXIV:0808.0500;%%
  
    \bibitem{Aharony}
  O.~Aharony, O.~Bergman, D.~L.~Jafferis and J.~Maldacena,
  ``N=6 superconformal Chern-Simons-matter theories, M2-branes and their gravity duals,''
  JHEP {\bf 0810} (2008) 091
  [arXiv:0806.1218 [hep-th]].
  %%CITATION = ARXIV:0806.1218;%%
  
  \bibitem{Seiberg}
  N.~Seiberg,
  ``Notes on theories with 16 supercharges,''
  Nucl.\ Phys.\ Proc.\ Suppl.\  {\bf 67} (1998) 158
  [hep-th/9705117].
  %%CITATION = HEP-TH/9705117;%%
  

   \bibitem{Gray}
   A.~Gray and L.~M.~Hervella,
   ``The sixteen classes of almost Hermitian manifolds and their linear invariants,''
   Ann. Mat. Pura Appl. {\bf 123} (1980) 35
  
  \bibitem{Nagy}
  P.~A.~Nagy, 
  ``NK geometry and Riemannian foliations,''
  Asian J. Math. {\bf 6} (2002) 481
  [math.DG/0203038].
  
  \bibitem{Grunewald}
  R.~Grunewald, 
  ``Six-dimensional Riemannian manifolds with a real Killing spinor,''
  Ann. Global Anal. Geom. {\bf 8} (1990) 43
  
    \bibitem{Bar}
  C.~B\"ar, 
  ``Real Killing spinors and holonomy,''
  Comm. Math. Phys. {\bf 154} (1993) 509
  
  \bibitem{Atiyah}
  M.~Atiyah and E.~Witten,
  ``M theory dynamics on a manifold of G(2) holonomy,''
  Adv.\ Theor.\ Math.\ Phys.\  {\bf 6} (2003) 1
  [hep-th/0107177].
  %%CITATION = HEP-TH/0107177;%%
  
  \bibitem{Acharya}
  B.~S.~Acharya and E.~Witten,
  ``Chiral fermions from manifolds of G(2) holonomy,''
  hep-th/0109152.
  %%CITATION = HEP-TH/0109152;%%
  
  \bibitem{Behrndt}
  K.~Behrndt and M.~Cvetic,
  ``General N = 1 supersymmetric flux vacua of (massive) type IIA string theory,''
  Phys.\ Rev.\ Lett.\  {\bf 95} (2005) 021601
  [hep-th/0403049].
  %%CITATION = HEP-TH/0403049;%%
  ``General N=1 supersymmetric fluxes in massive type IIA string theory,''
  Nucl.\ Phys.\ B {\bf 708} (2005) 45
  [hep-th/0407263].
  %%CITATION = HEP-TH/0407263;%%

 \bibitem{Chiossi}
  S.~Chiossi and S.~Salamon,
  ``The intrinsic torsion of $SU(3)$ and $G_2$ structures,''
  [math.DG/0202282].
  
    \bibitem{Verbitsky}
  M.~Verbitsky,
  ``An intrinsic volume functional on almost complex 6-manifolds and NK geometry,''
  Pacific \ J.\ Math. {\bf 235} (2008) 323.
  
  \bibitem{Boyer}
  C.~P.~Boyer and K.~Galicki
  ``Sasakian Geometry, Holonomy, and Supersymmetry,''
  [math.DG/0703231].
  
 \bibitem{Cvetic}
  M.~Cvetic, G.~W.~Gibbons, H.~Lu and C.~N.~Pope,
  ``Bianchi IX selfdual Einstein metrics and singular G(2) manifolds,''
  Class.\ Quant.\ Grav.\  {\bf 20} (2003) 4239
  [hep-th/0206151].
  %%CITATION = HEP-TH/0206151;%%
  
  \bibitem{Behrndt2}
  K.~Behrndt, G.~Dall'Agata, D.~Lust and S.~Mahapatra,
  ``Intersecting six-branes from new seven manifolds with G(2) holonomy,''
  JHEP {\bf 0208} (2002) 027
  [hep-th/0207117].
  %%CITATION = HEP-TH/0207117;%%
  
   \bibitem{Fernandez}
  M.~Fernandez, S.~Ivanov, V.~Mu\~noz and L.~Ugarte,
  ``Nearly hypo structures and compact NK 6-manifolds with conical singularities,''
  J. London Math. Soc. {\bf 78} (2008) 580.
  [math.DG/0602160].

   \bibitem{Butruille}
  J.~B.~Butruille,
  ``Homogeneous NK manifolds,''
  [math.DG/0612655].
 
  
  \bibitem{Loewy}
  A.~Loewy and Y.~Oz,
  ``Branes in special holonomy backgrounds,''
  Phys.\ Lett.\ B {\bf 537} (2002) 147
  [hep-th/0203092].
  %%CITATION = HEP-TH/0203092;%%
  


\bibitem{CasalderreySolana:2011us} 
  J.~Casalderrey-Solana, H.~Liu, D.~Mateos, K.~Rajagopal and U.~A.~Wiedemann,
  ``Gauge/String Duality, Hot QCD and Heavy Ion Collisions,'' CAmbridge University Press 2014, [arXiv:1101.0618 [hep-th]].  
  

  %\cite{Acharya:2003ii}
\bibitem{Acharya:2003ii}
  B.~S.~Acharya, F.~Denef, C.~Hofman and N.~Lambert,
  ``Freund-Rubin revisited,''
  hep-th/0308046.
  %%CITATION = HEP-TH/0308046;%%

  
  \bibitem{Benini}
  F.~Benini, F.~Canoura, S.~Cremonesi, C.~Nunez and A.~V.~Ramallo,
  ``Unquenched flavors in the Klebanov-Witten model,''
  JHEP {\bf 0702} (2007) 090
  [hep-th/0612118].
  %%CITATION = HEP-TH/0612118;%%
  
  \bibitem{Cassani}
  D.~Cassani and A.~-K.~Kashani-Poor,
  ``Exploiting N=2 in consistent coset reductions of type IIA,''
  Nucl.\ Phys.\ B {\bf 817} (2009) 25
  [arXiv:0901.4251 [hep-th]].
  %%CITATION = ARXIV:0901.4251;%%
  
  \bibitem{Haack}
  M.~Haack, D.~Lust, L.~Martucci and A.~Tomasiello,
  ``Domain walls from ten dimensions,''
  JHEP {\bf 0910} (2009) 089
  [arXiv:0905.1582 [hep-th]].
  %%CITATION = ARXIV:0905.1582;%%
  
  \bibitem{Gutowski:1999tu} 
  J.~Gutowski, G.~Papadopoulos and P.~K.~Townsend,
  ``Supersymmetry and generalized calibrations,''
  Phys.\ Rev.\ D {\bf 60}, 106006 (1999)
  [hep-th/9905156].
  
  %\cite{Koerber:2007hd}
\bibitem{Koerber:2007hd}
  P.~Koerber and D.~Tsimpis,
  ``Supersymmetric sources, integrability and generalized-structure compactifications,''
  JHEP {\bf 0708} (2007) 082
  [arXiv:0706.1244 [hep-th]].
  %%CITATION = ARXIV:0706.1244;%%
  
  \bibitem{Lust}
  D.~Lust and D.~Tsimpis,
  ``Supersymmetric AdS(4) compactifications of IIA supergravity,''
  JHEP {\bf 0502} (2005) 027
  [hep-th/0412250].
  %%CITATION = HEP-TH/0412250;%%
  
  \bibitem{Gaillard:2009kz} 
  J.~Gaillard and J.~Schmude,
  ``The Lift of type IIA supergravity with D6 sources: M-theory with torsion,''
  JHEP {\bf 1002}, 032 (2010)
  [arXiv:0908.0305 [hep-th]].

  
  \bibitem{Appelquist}
  T.~Appelquist and D.~Nash,
  ``Critical Behavior in (2+1)-dimensional {QCD},''
  Phys.\ Rev.\ Lett.\  {\bf 64} (1990) 721.
  %%CITATION = PRLTA,64,721;%%
  
  \bibitem{Polchinski:2010hw} 
  J.~Polchinski,
  ``Introduction to Gauge/Gravity Duality,''
  arXiv:1010.6134 [hep-th].
  
  %\cite{Niemi:1983rq}
\bibitem{Niemi:1983rq}
 A.~J.~Niemi and G.~W.~Semenoff,
 ``Axial Anomaly Induced Fermion Fractionization and Effective Gauge
Theory Actions in Odd Dimensional Space-Times,''
 Phys.\ Rev.\ Lett.\  {\bf 51}, 2077 (1983).
 %%CITATION = PRLTA,51,2077;%%
  
  \bibitem{Redlich}
  A.~N.~Redlich,
  ``Gauge Noninvariance and Parity Violation of Three-Dimensional Fermions,''
  Phys.\ Rev.\ Lett.\  {\bf 52} (1984) 18.
  %%CITATION = PRLTA,52,18;%%
  ``Parity Violation and Gauge Noninvariance of the Effective Gauge Field Action in Three-Dimensions,''
  Phys.\ Rev.\ D {\bf 29} (1984) 2366.
  %%CITATION = PHRVA,D29,2366;%%
  
 \bibitem{Jafferis:2011zi} 
  D.~L.~Jafferis, I.~R.~Klebanov, S.~S.~Pufu and B.~R.~Safdi,
  ``Towards the F-Theorem: N=2 Field Theories on the Three-Sphere,''
  JHEP {\bf 1106}, 102 (2011)
  [arXiv:1103.1181 [hep-th]].
  %%CITATION = ARXIV:1103.1181;%%

  
  \bibitem{Gutowski}
  J.~Gutowski, S.~Ivanov and G.~Papadopoulos,
  ``Deformations of generalized calibrations and compact nonKahler manifolds with vanishing first Chern class,''
  Asian J. Math. {\bf 7} (2003), 39
   [math.DG/0205012]
  %%CITATION = MATH/0205012;%%
  
  
  \bibitem{Koerber} 
  P.~Koerber and L.~Martucci,
  ``D-branes on AdS flux compactifications,''
  JHEP {\bf 0801} (2008) 047
  [arXiv:0710.5530 [hep-th]].
  %%CITATION = ARXIV:0710.5530;%%
  

  
\bibitem{Karch:2002sh} 
  A.~Karch and E.~Katz,
  ``Adding flavor to AdS / CFT,''
  JHEP {\bf 0206}, 043 (2002)
  [hep-th/0205236].  

\bibitem{Kruczenski:2003be} 
  M.~Kruczenski, D.~Mateos, R.~C.~Myers and D.~J.~Winters,
  ``Meson spectroscopy in AdS / CFT with flavor,''
  JHEP {\bf 0307}, 049 (2003)
  [hep-th/0304032].


\bibitem{Babington:2003vm} 
  J.~Babington, J.~Erdmenger, N.~J.~Evans, Z.~Guralnik and I.~Kirsch,
  ``Chiral symmetry breaking and pions in nonsupersymmetric gauge / gravity duals,''
  Phys.\ Rev.\ D {\bf 69}, 066007 (2004)
  [hep-th/0306018].

\bibitem{Mateos:2006nu} 
  D.~Mateos, R.~C.~Myers and R.~M.~Thomson,
  ``Holographic phase transitions with fundamental matter,''
  Phys.\ Rev.\ Lett.\  {\bf 97}, 091601 (2006)
  [hep-th/0605046].

\bibitem{Mateos:2007vn} 
  D.~Mateos, R.~C.~Myers and R.~M.~Thomson,
  ``Thermodynamics of the brane,''
  JHEP {\bf 0705}, 067 (2007)
  [hep-th/0701132].
  
  \bibitem{prep} 
  A.~F.~Faedo, A.~Kundu, D.~Mateos, C.~Pantelidou and J.~Tarrio,
  ``Super Yang-Mills with Compressible Quark Matter,'' to appear.
  
\bibitem{Faedo:2014ana} 
  A.~F.~Faedo, A.~Kundu, D.~Mateos and J.~Tarrio,
  ``(Super)Yang-Mills at Finite Heavy-Quark Density,''
  JHEP {\bf 1502}, 010 (2015)
  [arXiv:1410.4466 [hep-th]].

  \bibitem{Kashani}
  A.~-K.~Kashani-Poor,
  ``Nearly Kaehler Reduction,''
  JHEP {\bf 0711} (2007) 026
  [arXiv:0709.4482 [hep-th]].
  %%CITATION = ARXIV:0709.4482;%%


\end{thebibliography}
\end{document}